\documentclass[prb,twocolumn,amsmath,amssymb]{revtex4}
\usepackage{graphicx}

\begin{document}


\title{Connecting Berry's phase and the pumped charge in
a Cooper pair pump}


\author{M. Aunola}
\email[email: ]{Matias.Aunola@phys.jyu.fi}
\author{J.~J. Toppari}
\email[email: ]{Jussi.Toppari@phys.jyu.fi}
\affiliation{Department of Physics, University of Jyv\"askyl\"a,
P.O. Box 35 (YFL), FIN-40014 University of Jyv\"askyl\"a, FINLAND}

\date{\today}

\begin{abstract}

The properties of the tunnelling-charging Hamiltonian of a Cooper pair
pump are well understood in the regime of weak and intermediate
Josephson coupling, i.e.  when $E_{\mathrm{J}}\lesssim E_{\mathrm{C}}$.
It is also known that  Berry's phase is related to 
the pumped charge induced by the adiabatical variation
of the eigenstates. We show explicitly that pumped charge in Cooper pair pump
can be understood as a partial derivative of Berry's phase
with respect to the phase difference $\phi$ across the array.
The phase fluctuations always present in real experiments can also be 
taken into account, although only approximately. Thus the measurement of 
the pumped current gives reliable, yet indirect, information on Berry's phase. 
As closing remarks, we give the differential relation between Berry's phase 
and the pumped charge, and state that the mathematical results are valid for any 
observable expressible as a partial derivative of the Hamiltonian.

\end{abstract}


\maketitle


Josephson junction devices, e.g.~Cooper pair box, superconducting
single electron transistor (SSET) and Cooper pair pump (CPP), have
been extensively studied both
theoretically\cite{ave98,pek99,mak99,ave00,fal00,pek01} and
experimentally.\cite{eil94,moo99,nak99,orl99,cho01,bib02} (For a recent
review, see Ref.~\onlinecite{mak01}.)  Possible applications include
coherent Cooper pair pumping\cite{pek99} with related  decoherence 
studies\cite{nak99,pek01} or metrological
applications,\cite{has99} and
the use of these devices as superconducting quantum bits (squbits) in
quantum computation.\cite{ave98,moo99,zhu02} In this paper we focus on CPP
whose idealised tunnelling-charging Hamiltonian has been studied in
detail in Refs.~\onlinecite{pek99,aun00,aun01}.  For closed loops in the parameter 
space, we relate the pumped charge to Berry's phase, a well-known geometrical 
phase attained by an 
adiabatically evolving eigenstate of a time-dependent
Hamiltonian.\cite{ber84,sim83,nak90} Some applications of geometrical
phases in mesoscopic systems, are discussed in 
Refs.~\onlinecite{fal00,cha95} and the references therein. 
We illustrate the results both for the SSET
and a CPP, and consider the effects due to phase fluctuations, present
when experimentally measuring the pumped current.

In a CPP the pumping of Cooper pairs is induced by cyclic variation of the gate
voltages while the evolution of the total phase difference across the array, $\phi$,
is either fixed by ideal biasing\cite{pek99} or stochastically
decoherent. Theoretical predictions are based on the adiabatic
evolution of the eigenstates which splits the induced current into two
parts: \cite{pek99,aun00} The direct supercurrent, which flows constantly and
is proportional to the $\phi$-derivative of the dynamical phase of
the eigenstate. The other part, the pumped charge, is explicitly
induced by the action of pumping and proportional to the
$\phi$-derivative of Berry's phase for closed loops. Existence of such a relation was
already implicitly stated in Ref.~\onlinecite{pek99}.
The underlying reason for these connections is that the supercurrent operator
$I_{\mathrm{S}}$ is an operator derivative\cite{rud73}
of the full Hamiltonian with respect to $\phi$. This also implies that
all of the results obtained in this paper are valid for any
observable expressible as a partial derivative of the 
corresponding Hamiltonian. However, in real applications it might be 
reasonable to use the nonadiabatically
attained geometrical phase instead of Berry's phase.\cite{zhu02}

\begin{figure}[htb]
\includegraphics[width=54truemm]{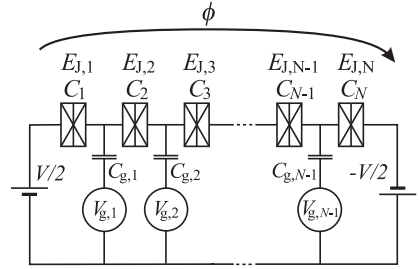}
\caption{An ideal superconducting array of 
Josephson junctions (CPP). Here $C_k$ and $E_{\mathrm{J},k}$ are the
capacitance and the Josephson energy of the $k^{\mathrm{th}}$ junction,
respectively. The total phase difference across the array, $\phi$,
is a constant of motion in the absence of the external 
bias voltage $V$.
\label{fig:pump}}
\end{figure}

A schematic view of a CPP is shown in
Fig.~\ref{fig:pump}. We assume that the gate voltages
$V_{\mathrm{g},j}$ are independent and externally operated.
The ideally operated bias voltage across the array, 
$V$, controls the total phase difference, $\phi$,
according to  $d\phi/dt=-2eV/\hbar$. In the absence of 
bias voltage,  $\phi$ remains fixed and 
becomes a good quantum number.\cite{ing92,pek99}
Conversely, the conjugate variable $\hat M$,  
the average number of tunnelled Cooper pairs, 
becomes completely undetermined.

The tunnelling-charging Hamiltonian
\begin{equation}
H=H_{\mathrm{C}}(\vec q\,)+H_{\mathrm{J}},\label{eq:simphami}
\end{equation}
is assumed to be the correct description of the microscopic system,
neglecting quasiparticle tunnelling
as well as other degrees of freedom. The charging Hamiltonian 
$H_{\mathrm{C}}(\vec q\,)$ depends on the normalised gate charges,
$\vec q:=(q_1,\ldots,q_{N-1})$,  and 
the  number of Cooper pairs on each island, 
$\vec n=(n_1,\ldots,n_{N-1})$, according to
$\langle \vec n\vert H_{\rm C}(\vec q\,)\vert \vec n'\rangle_{\phi}=
E_{\rm C}(\vec n-\vec q\,)\delta_{\vec n,\vec n'}$.
The function $E_{\rm C}(\vec x)$ gives the details of the charging
energy, see e.g. Ref.~\onlinecite{aun00}.
The Josephson (tunnelling) Hamiltonian is given by 
\begin{equation}
H_{\mathrm{J}}=-\sum_{k=1}^N E_{\mathrm{J},k}\cos(\phi_k),
\label{eq:tunnelit}
\end{equation}
where $E_{\mathrm{J},k}$ is the Josephson coupling energy of the
$k^{\mathrm{th}}$ junction. The average supercurrent operator 
can be written in the form
\begin{equation}
I_{\mathrm{S}}=\frac{-2e}{N\hbar}\sum_{k=1}^N E_{\mathrm{J},k}\sin(\phi_k)
=\frac{-2e}{\hbar}\frac{\partial H}{\partial\phi},\label{eq:superc}
\end{equation}
where the operator derivative\cite{rud73} 
is defined as
\begin{equation}
\frac{\partial H}{\partial\phi}:=
\lim_{\phi'\rightarrow \phi}\frac{H(\phi')-H(\phi)}{\phi'-\phi}.
\end{equation}
In the preferred representation, the  parameter space is an 
$N$-dimensional manifold $\mathbb{R}^{N-1}\times S^{1}$, where
the elements are of the form 
$\vec p=(\vec q,\phi)$ with $\phi\in[0,2N\pi)$.
For $N=2$, the system reduces to  a SSET
whose Hamiltonian is discussed
in Ref.~\onlinecite{tink96}. Asymptotically
exact eigenvectors for strong Josephson coupling
are given in Ref.~\onlinecite{aun02d}.

We will study the adiabatic evolution of instantaneous energy 
eigenstates $\vert m\rangle$, while 
changing the gate voltages along a closed path 
$\Gamma:t\mapsto \vec p^{\,}(t)=(\vec q^{\,}(t),\phi(t))$ with
$t\in [0,\tau]$. This induces  a charge transfer
$Q_{\mathrm{tot}}:=Q_{\mathrm{s}}+Q_{\mathrm{p}}$, where
the pumped charge, $Q_{\mathrm{p}}$, depends only on
the chosen path, $\Gamma$, determined by the gating sequence, while the charge 
transferred by the direct supercurrent, 
$Q_{\mathrm{s}}$, also depends on the rate of operation of the 
gate voltages, i.e.,~the value of $\tau$. The total transferred 
charge, in units of $-2e$, for state $\vert m\rangle$ 
becomes\cite{pek99,aun01}
\begin{eqnarray}
Q_{\mathrm{tot}}=-\frac{\partial \eta_m}{\partial \phi}+
2\oint_{\Gamma}{\rm Re}\left[\langle m\vert\hat M
\vert dm\rangle\right].
\label{eq:simppump}
\end{eqnarray}
Here $\hat M=-i\partial/\partial\phi$ is the operator for average number of tunnelled Cooper
pairs,  $\eta_m=-\int_0^\tau (E_m(t)/\hbar)dt$
is the dynamical phase and
$\vert dm\rangle$ is the
change in the eigenstate $\vert m\rangle$ due to a differential change 
$d\vec p=(d\vec q,d\phi)$. 
A change $d\phi$ in the phase
difference at $\vec p$  induces no pumped charge as we find
$dQ_p=2{\rm Im}[\langle \hat M m\vert
\hat M m\rangle] d\phi=0$.
In other words, the bias voltage $V$ induces no
pumped charge for fixed gate voltages. 

The expression for $Q_{\mathrm{p}}$ is rather similar to
the corresponding  Berry's phase\cite{ber84,sim83}
\begin{equation}
\gamma_m(\Gamma)=i\oint_{\Gamma} \langle m\vert dm\rangle.
\label{eq:berry}
\end{equation}
It should be stressed that Eqs.~(\ref{eq:simppump}) and (\ref{eq:berry})
are well defined also for open paths.
The derivative $d$ in Eq.~(\ref{eq:berry}) is an exterior derivative so, for a closed path $\Gamma$,
we may integrate Berry's curvature over a two-surface\cite{nak90}
\begin{equation}
\gamma_m(\Gamma)=i\sum_{k=1}^N\sum_{l=1}^N\iint_{S_\Gamma}
\frac{\partial\langle m\vert}{\partial q_k}
\frac{\partial\vert m\rangle}{\partial q_l} dq_{k}\wedge dq_l,
\label{eq:berry2}
\end{equation}
where the boundary of $S_\Gamma$ is the path $\Gamma$.\cite{integr}

We now construct an extended path for which Berry's phase 
is proportional to the charge pumped along the path $\Gamma$.
Let us define a class of closed paths
$\{\Gamma_\varphi\}$ by
$\Gamma_\varphi:t\mapsto \vec p^{\,}(t)=(\vec q^{\,}(t),
\phi(t)+\varphi)$, where $t\in [0,\tau]$, so that $\Gamma_0=\Gamma$.
The inverse of a path is the same path traversed in the
opposite direction, which also holds for paths with distinct
end points. We define an additional class of paths $\{\varphi_{\vec p}\}$
according to
$\varphi_{\vec p}:
t\mapsto (\vec q,\phi+t\varphi)$, where
$t\in[0,1]$.  The  extended path 
\begin{equation}
\Gamma_{\mathrm{ext}}^{\varphi}:=\Gamma_0\circ \varphi_{\vec p\,(0)}
\circ
\Gamma_\varphi^{-1}\circ \varphi_{\vec p\,(0)}^{-1}
\end{equation}
is also closed and spans  a two-dimensional 
integration surface
whose width in $\phi$-direction is $\varphi$. 
By traversing the boundary 
the contributions from $\varphi_{\vec p\,(0)}$ and 
$\varphi_{\vec p\,(0)}^{-1}$ naturally cancel, and we find
\begin{equation}
\gamma(\Gamma_{\mathrm{ext}}^{\varphi})=\gamma(\Gamma)-
\gamma(\Gamma_\varphi).\label{eq:berry33}
\end{equation}

\begin{figure}[htb]
\includegraphics[width=42truemm]{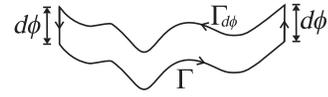}
\caption{A closed path $\Gamma$ has been flattened, i.e.~the
 ends of $\Gamma$ meet. The infinitesimal pieces of length 
$d\phi$ are identical, but traversed in opposite directions.
The two-dimensional integration surface $S_{\Gamma,d\phi}$ 
is spanned between orbits $\Gamma$ and $\Gamma_{d\phi}$. 
Berry's phase corresponding to the boundary of $S_{\Gamma,d\phi}$ 
is identical to $-Q_{\mathrm{p}}(\Gamma)d\phi$, with 
$Q_{\mathrm{p}}(\Gamma)$ defined in 
Eq.~(\ref{eq:pumpedberry}).
\label{fig:strip}}
\end{figure}

Next we take the limit $\varphi\rightarrow 0$ and 
consider a strip of infinitesimal width $d\phi$ between $\Gamma_{d\phi}$ 
and $\Gamma_0$ as illustrated in
Fig.~\ref{fig:strip}.
This means that in Eq.~(\ref{eq:berry2}) we have either 
$dq_k=d\phi$ or $dq_l=d\phi$ as the full length of integration and
 we can factor $d\phi$ from the expression for Berry's phase.
By rephrasing $Q_{\mathrm{p}}$ in Eq.~(\ref{eq:simppump}) as
\begin{equation}
Q_{\mathrm{p}}(\Gamma)=i\sum_{k=1}^N
\oint_{\Gamma}\left[\frac{\partial\langle 
m\vert}{\partial \phi}\frac{\partial\vert m\rangle}{\partial q_k}
-\frac{\partial\langle m\vert}{\partial q_k}
\frac{\partial\vert m\rangle}{\partial \phi}
\right]dq_k,
\label{eq:pumpedberry}
\end{equation}
we see that it is identical to Berry's phase in Eq.~(\ref{eq:berry2}) apart 
from the factor $-d\phi$. By taking the limit from
the equivalent result  $\lim_{\varphi\rightarrow 0}
(\gamma(\Gamma)-\gamma(\Gamma_\varphi))/\varphi$,
we obtain the first main result
\begin{equation}
Q_{\mathrm{s}}+Q_{\mathrm{p}}=-\partial\eta_m/\partial \phi-
\partial\gamma_m/\partial \phi.
\label{qp}
\end{equation}
This clearly shows the connection between Berry's phase and 
the pumped charge, which is completely analogous with 
the connection between the dynamical phase and the
accumulated charge due to direct supercurrent. 

\begin{figure}[htb]
\includegraphics[width=40truemm]{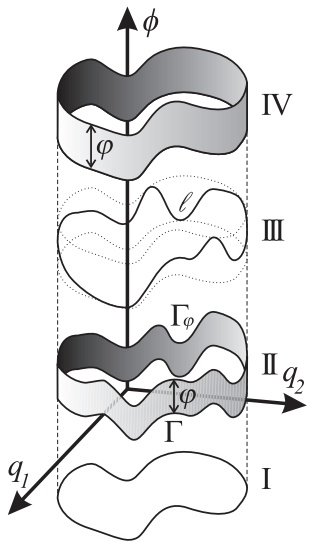}
\caption{I) A projection of path $\Gamma$ in the $(q_1,q_2)$-plane. II) A strip of finite
width $\varphi$ based on the path $\Gamma$.  III) The fluctuations of
$\phi$ on a single pumping cycle, $\ell$. IV) Ideal operation of gate voltages
produces a strip bounded by the planes $\phi=\phi_0$ and
$\phi=\phi_0+\varphi$. The same result is obtained approximately after
many cycles with restricted, stochastic fluctuations of $\phi$.}
\label{fig:orbits}
\end{figure}

We now proceed in the opposite direction and 
consider strips of finite width instead of infinitesimal ones.
By integrating the pumped charge with respect to $\phi$ over the set  
$\{\Gamma_\varphi\}$, we obtain
the average pumped charge per cycle, $Q_{p,\mathrm{ave}}$, 
as
\begin{equation}
Q_{p,\mathrm{ave}}=
\frac1{\varphi}\int_0^\varphi Q_p(\Gamma_\phi)d\phi=
\frac{\gamma(\Gamma_0)-\gamma(\Gamma_\varphi)}{\varphi}.
\label{eq:qave}
\end{equation}
The graphical representation of this situation in 
a three-junction CPP and a SSET, 
are shown in Fig.~\ref{fig:orbits} (I,II) and in Fig.~4, respectively. 
The cases are qualitatively different,
because there is only one $q$-coordinate in a SSET.

\begin{figure}[htb]
\includegraphics[width=45truemm]{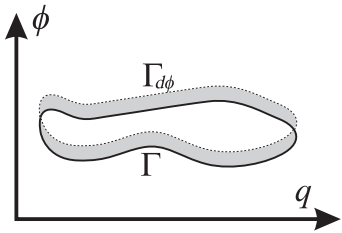}
\caption{Pumping of Cooper pairs in SSET
recuires also an ideal control of $\phi$. The projection
of the path $\Gamma$ onto $q$-space is a line traversed
back and forth.}
\label{fig:SSET}
\end{figure}

We now wish to relate the above results to actual 
measurements of Cooper pair pumping. First, consider a closed
path $\Gamma$ corresponding to a fixed value of $\phi_0$
as in Ref.~\onlinecite{pek99}. Under ideal operation  of gate
and bias voltages we can change $\phi$ slightly 
between each cycle and obtain a clean strip bounded by
the planes $\phi=\phi_0$ and $\phi=\phi_0+\varphi$ as shown
in Fig.~\ref{fig:orbits} (IV). Combined with Eq.~(\ref{eq:qave})
this amounts to an important result  
for an ideal CPP: The measured
pumped charge per cycle (i.e.~ $Q_{p,\mathrm{ave}}$) yields direct
information about differences of Berry's phases. 

Obtaining the same information in a real experiment is not so
straightforward. Neither the phase difference, $\phi$, nor the gate
voltages are ideally controlled. Nevertheless, we try to
partially circumvent these problems using reasonable 
approximations. First, we assume that the gate voltages
are operated accurately enough, so that the projections
onto $q$-space nearly coincide. 
Additionally, $\phi$ fluctuates stochastically,
but these fluctuations are restricted during time intervals shorter
than the decoherence time, $\tau_\phi$.\cite{pek01} 
For times larger than $\tau_\phi$ 
the fluctuations mount up too large and 
the phase coherence of the system is lost.
  
The decoherence is induced by any interaction between 
the quantum mechanical system and the environment. 
In a well prepared experiment,
e.g.~a CPP can be isolated from its surroundings
so that the main contribution to $\tau_\phi$ is
given by the electromagnetic environment in the vicinity of the sample
and the effects due to finite temperature, restricting the
measurements to subkelvin regime.

The decoherence time can be calculated theoretically from the fluctuation-dissipation
--theorem as in Ref.~\onlinecite{pek01} or by looking at coherences, i.e., 
off-diagonal elements of the density matrix. The Hamiltonian in 
the presence of the electromagnetic environment reads
\begin{equation}
H = H_{\mathrm{C}}(q) +  H_{\mathrm{J}} + H_{\mathrm{env}} + H_{\mathrm{int}},
\end{equation}	
where $H_{\mathrm{env}} = \sum_j(b_j^\dag b_j + 1/2)\hbar\omega_j$ 
and $b_j^\dag$ and $b_j$ are the creation and annihilation 
operators of the bosonic environmental mode $j$ with energy $\hbar
\omega_j$, respectively.\cite{cald83} As an example, we consider
a SSET but it should be stressed that the result generalises for 
any number of junctions. We write the density 
matrix $\rho^{\vec k} = \Psi^{\vec k}\Psi^{\vec k \dag}$
in the basis consisting of two SSET 
charge states, $\{|m\rangle\}_{m=0}^1$  and environmental modes
$\{|\vec k= (k_1, k_2, \ldots) \rangle\}$.
Then the Hamiltonian describing the interaction between 
SSET and the environment becomes \cite{cald83,cottet2000}
\begin{equation}
H_{\mathrm{int}} = -i\sqrt{\pi}\sum
\limits_j\hbar\omega_j \sqrt{\frac{Z_j}{R_{\mathrm{K}}}}
\left(b_j - b_j^\dag\right)\left(
\begin{array}{cc}
1 &0 \\ 0&-1\end{array}\right),
\end{equation}
where $Z_j$ is the impedance of the mode $j$ and
$R_{\mathrm{K}} = h/e^2 \simeq  25.8$ k$\Omega$ is the resistance quantum. 

The equation of motion for $\rho^{\vec k}_{\rm I}$
in the interaction 
picture is given by the Liouville equation, 
$i\hbar (d\rho^{\vec k}_{\mathrm{I}}(t)/dt) = [H_{\mathrm{int, I}}\,,
\rho^{\vec k}_{\mathrm{I}}(t)]$.
By solving the differential equation for the coherence
matrix elements and tracing out the environmental 
configurations $\vec k$ we obtain the final result
\begin{equation}
\rho_{\mathrm{I}, 12}(t) = \rho_{\mathrm{I}, 12}(0)
\exp\left[-2{\rm Re} J(t)\right],
\label{Jt}
\end{equation}
 which corresponds to the same time scale as given by the fluctuation-dissipation
-6-theorem. \cite{pek01}
Here $J(t)$ is the phase-phase correlation function $J(t) = 
\langle\left[\varphi(t)-\varphi(0)\right]\varphi(0)\rangle$ 
and $\rho_{\mathrm{I}} = \langle\rho_{\mathrm{I}}^{\vec k}\rangle_{\vec k}\,$.\cite{ing92}
In case of purely resistive electromagnetic environment, 
$R_{\mathrm{e}}$, Eq.~(\ref{Jt}) yields \cite{pek01}
$\tau_\phi \simeq [\pi\hbar/(16k_{\mathrm{B}}T)]R_{\mathrm{K}}/R_{\mathrm{e}}$, 
where we have assumed nonzero temperature and $\pi k_{\mathrm{B}}Tt/\hbar\gg 1$.
For realistic measurement parameters, e.g. 
$T= 10$ mK and $R_{\mathrm{e}}=10$ $\Omega$, one
obtains a rather long time $\tau_\phi \simeq 0.4$ $\mu$s.

Returning to Berry's phase, we assume an initial value $\phi_0$ and consider
time intervals shorter than $\tau_\phi$, effectively restricting
$\phi$ to a finite range $[\phi_1,\phi_2]\ni\phi_0$. If sufficiently many
(identical) cycles of gate voltages are performed during this time,
the fluctuations of $\phi$ yield a relatively thick mesh of 
trajectories within the strip. Although the 'weights' for different
values of $\phi$ are uneven, we approximate the mesh with a
uniform distribution which is a subset of the range $[\phi_1,\phi_2]$.
This corresponds to a well-defined strip as in the ideal case of Eq.~(\ref{eq:qave}) 
and is presented in Fig.~\ref{fig:orbits} (IV).
A cycle, $\ell$, with exaggerated fluctuations in $\phi$, is shown in 
Fig.~\ref{fig:orbits} (III).
Due to the stochastic nature of the fluctuations, it is impossible
to \emph{predict} the correct range to be used. Nevertheless, for
periods that are short enough, the correspondence between Berry's phase
and the measured pumped charge exists in the sense of  Eq.~(\ref{eq:qave}). 
If the end points of the full pumping cycle are sufficiently close,
Eq.~(\ref{qp}) is valid, at least in the framework of the model.

\begin{figure}[htb]
\includegraphics[width=60truemm]{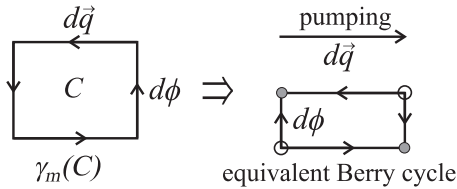}
\caption{{}An infinitesimal cycle $C$ corresponding to Berry's
phase $\gamma_m(C)$ consists of four legs. The charge transfer
$Q_{\rm p}$ for a fixed $\phi$ is identical to Berry's
phase induced by traversing the legs in the directions 
shown on the right-hand-side divided by $d\phi$.
This path can not be continuously followed in the
parameter space.}  \label{fig:cycle}
\end{figure}

As a final note, we construct the differential relation between
Berry's phase and the pumped charge. Let us
consider Berry's phase $\gamma_{m}$ induced by an 
infinitesimal closed cycle $C$ at $(\vec q,\phi)$ with 
sides $d\vec q$ and $d\phi$ as shown by the left-hand-side of 
Fig.~\ref{fig:cycle}.
On the right-hand-side, the pumped charge due to $d\vec q$
multiplied by $d\phi$, is identical
to Berry's phase induced by the discontinuous path below it.
By following any closed pumping path $\Gamma$ and 
integrating the pumped charge, we recover Eq.~(\ref{qp}).
If the path is not a closed one, a nontrivial integration with 
respect  to $\phi$ remains, regardless of the width of
the strip. 

In conclusion, we have shown explicitly how the pumped 
charge in Cooper pair pump can be understood as a partial 
derivative of Berry's phase with respect to the phase difference 
$\phi$ across the array. We have only used the
fact that the supercurrent operator $I_{\mathrm{S}}$ is 
an operator derivative of the full Hamiltonian. 
Thus these results generalise for any observable
with this property. We have also shown 
how one could obtain information about Berry's 
phase by measuring the pumped current in a CPP.

\begin{acknowledgments}
This work has been supported by the Academy of Finland
under the Finnish Centre of Excellence Programme 2000-2005
(Project No. 44875, Nuclear and Condensed Matter Programme at JYFL).
The authors thank Dr.~K.~Hansen, Dr.~A.~Cottet and Prof. 
J.~P.~Pekola for discussions and comments.
\end{acknowledgments}

\bibliography{Largeejc}

\begin{thebibliography}{28}
\expandafter\ifx\csname natexlab\endcsname\relax\def\natexlab#1{#1}\fi
\expandafter\ifx\csname bibnamefont\endcsname\relax
  \def\bibnamefont#1{#1}\fi
\expandafter\ifx\csname bibfnamefont\endcsname\relax
  \def\bibfnamefont#1{#1}\fi
\expandafter\ifx\csname citenamefont\endcsname\relax
  \def\citenamefont#1{#1}\fi
\expandafter\ifx\csname url\endcsname\relax
  \def\url#1{\texttt{#1}}\fi
\expandafter\ifx\csname urlprefix\endcsname\relax\def\urlprefix{URL }\fi
\providecommand{\bibinfo}[2]{#2}
\providecommand{\eprint}[2][]{\url{#2}}

\bibitem[{\citenamefont{Averin}(1998)}]{ave98}
\bibinfo{author}{\bibfnamefont{D.~V.} \bibnamefont{Averin}},
  \bibinfo{journal}{Solid State Commun.} \textbf{\bibinfo{volume}{105}},
  \bibinfo{pages}{659} (\bibinfo{year}{1998}).

\bibitem[{\citenamefont{Pekola et~al.}(1999)\citenamefont{Pekola, Toppari,
  Savolainen, and Averin}}]{pek99}
\bibinfo{author}{\bibfnamefont{J.~P.} \bibnamefont{Pekola}},
  \bibinfo{author}{\bibfnamefont{J.~J.} \bibnamefont{Toppari}},
  \bibinfo{author}{\bibfnamefont{M.~T.} \bibnamefont{Savolainen}},
  \bibnamefont{and} \bibinfo{author}{\bibfnamefont{D.~V.}
  \bibnamefont{Averin}}, \bibinfo{journal}{Phys.~Rev.~B}
  \textbf{\bibinfo{volume}{60}}, \bibinfo{pages}{R9931} (\bibinfo{year}{1999}).

\bibitem[{\citenamefont{Makhlin et~al.}(1999)\citenamefont{Makhlin,
  Sch$\ddot{\mathrm{o}}$n, and Shnirman}}]{mak99}
\bibinfo{author}{\bibfnamefont{Y.}~\bibnamefont{Makhlin}},
  \bibinfo{author}{\bibfnamefont{G.}~\bibnamefont{Sch$\ddot{\mathrm{o}}$n}},
  \bibnamefont{and} \bibinfo{author}{\bibfnamefont{A.}~\bibnamefont{Shnirman}},
  \bibinfo{journal}{Nature} \textbf{\bibinfo{volume}{386}},
  \bibinfo{pages}{305} (\bibinfo{year}{1999}).

\bibitem[{\citenamefont{Averin}(2000)}]{ave00}
\bibinfo{author}{\bibfnamefont{D.~V.} \bibnamefont{Averin}}, in
  \emph{\bibinfo{booktitle}{Exploring the quantum-classical frontier}}, edited
  by \bibinfo{editor}{\bibfnamefont{J.~R.} \bibnamefont{Friedman}}
  \bibnamefont{and} \bibinfo{editor}{\bibfnamefont{S.}~\bibnamefont{Han}}
  (\bibinfo{publisher}{Nova science publishers}, \bibinfo{address}{Commack,
  NY}, \bibinfo{year}{2000}), \bibinfo{note}{cond-mat/0004364}.

\bibitem[{\citenamefont{Falci et~al.}(2000)\citenamefont{Falci, Fazio, Palma,
  Siewert, and Vedral}}]{fal00}
\bibinfo{author}{\bibfnamefont{G.}~\bibnamefont{Falci}},
  \bibinfo{author}{\bibfnamefont{R.}~\bibnamefont{Fazio}},
  \bibinfo{author}{\bibfnamefont{G.~H.} \bibnamefont{Palma}},
  \bibinfo{author}{\bibfnamefont{J.}~\bibnamefont{Siewert}}, \bibnamefont{and}
  \bibinfo{author}{\bibfnamefont{V.}~\bibnamefont{Vedral}},
  \bibinfo{journal}{Nature} \textbf{\bibinfo{volume}{407}},
  \bibinfo{pages}{355} (\bibinfo{year}{2000}).

\bibitem[{\citenamefont{Pekola and Toppari}(2001)}]{pek01}
\bibinfo{author}{\bibfnamefont{J.~P.} \bibnamefont{Pekola}} \bibnamefont{and}
  \bibinfo{author}{\bibfnamefont{J.~J.} \bibnamefont{Toppari}},
  \bibinfo{journal}{Phys.~Rev.~B} \textbf{\bibinfo{volume}{64}},
  \bibinfo{pages}{172509} (\bibinfo{year}{2001}).

\bibitem[{\citenamefont{Eiles and Martinis}(1994)}]{eil94}
\bibinfo{author}{\bibfnamefont{T.~M.} \bibnamefont{Eiles}} \bibnamefont{and}
  \bibinfo{author}{\bibfnamefont{J.~M.} \bibnamefont{Martinis}},
  \bibinfo{journal}{Phys.~Rev. B} \textbf{\bibinfo{volume}{64}},
  \bibinfo{pages}{R627} (\bibinfo{year}{1994}).

\bibitem[{\citenamefont{Mooij et~al.}(1999)\citenamefont{Mooij, Orlando,
  Levitov, Tian, van~der Val, and Lloyd}}]{moo99}
\bibinfo{author}{\bibfnamefont{J.~E.} \bibnamefont{Mooij}},
  \bibinfo{author}{\bibfnamefont{T.~P.} \bibnamefont{Orlando}},
  \bibinfo{author}{\bibfnamefont{L.}~\bibnamefont{Levitov}},
  \bibinfo{author}{\bibfnamefont{L.}~\bibnamefont{Tian}},
  \bibinfo{author}{\bibfnamefont{C.~H.} \bibnamefont{van~der Val}},
  \bibnamefont{and} \bibinfo{author}{\bibfnamefont{S.}~\bibnamefont{Lloyd}},
  \bibinfo{journal}{Science} \textbf{\bibinfo{volume}{285}},
  \bibinfo{pages}{1036} (\bibinfo{year}{1999}).

\bibitem[{\citenamefont{Nakamura et~al.}(1999)\citenamefont{Nakamura, Pashkin,
  and Tsai}}]{nak99}
\bibinfo{author}{\bibfnamefont{Y.}~\bibnamefont{Nakamura}},
  \bibinfo{author}{\bibfnamefont{Y.~A.} \bibnamefont{Pashkin}},
  \bibnamefont{and} \bibinfo{author}{\bibfnamefont{J.~S.} \bibnamefont{Tsai}},
  \bibinfo{journal}{Nature} \textbf{\bibinfo{volume}{398}},
  \bibinfo{pages}{786} (\bibinfo{year}{1999}).

\bibitem[{\citenamefont{Orlando et~al.}(1999)\citenamefont{Orlando, Mooij,
  Tian, van~der Val L.~Levitov, , and Lloyd}}]{orl99}
\bibinfo{author}{\bibfnamefont{T.~P.} \bibnamefont{Orlando}},
  \bibinfo{author}{\bibfnamefont{J.~E.} \bibnamefont{Mooij}},
  \bibinfo{author}{\bibfnamefont{L.}~\bibnamefont{Tian}},
  \bibinfo{author}{\bibfnamefont{C.~H.} \bibnamefont{van~der Val L.~Levitov}},
  , \bibnamefont{and} \bibinfo{author}{\bibfnamefont{S.}~\bibnamefont{Lloyd}},
  \bibinfo{journal}{Phys.~Rev.~B.} \textbf{\bibinfo{volume}{60}},
  \bibinfo{pages}{15398} (\bibinfo{year}{1999}).

\bibitem[{\citenamefont{Choi et~al.}(2001)\citenamefont{Choi, Fazio, Siewert,
  and Bruder}}]{cho01}
\bibinfo{author}{\bibfnamefont{M.~S.} \bibnamefont{Choi}},
  \bibinfo{author}{\bibfnamefont{R.}~\bibnamefont{Fazio}},
  \bibinfo{author}{\bibfnamefont{J.}~\bibnamefont{Siewert}}, \bibnamefont{and}
  \bibinfo{author}{\bibfnamefont{C.}~\bibnamefont{Bruder}},
  \bibinfo{journal}{Europhys. Lett.} \textbf{\bibinfo{volume}{53}},
  \bibinfo{pages}{251} (\bibinfo{year}{2001}).

\bibitem[{\citenamefont{Bibow et~al.}(2002)\citenamefont{Bibow, Lafarge, and
  Levy}}]{bib02}
\bibinfo{author}{\bibfnamefont{E.}~\bibnamefont{Bibow}},
  \bibinfo{author}{\bibfnamefont{P.}~\bibnamefont{Lafarge}}, \bibnamefont{and}
  \bibinfo{author}{\bibfnamefont{L.~P.} \bibnamefont{Levy}},
  \bibinfo{journal}{Phys.~Rev.~Lett} \textbf{\bibinfo{volume}{88}},
  \bibinfo{pages}{017003} (\bibinfo{year}{2002}).

\bibitem[{\citenamefont{Makhlin et~al.}(2001)\citenamefont{Makhlin,
  Sch$\ddot{\mathrm{o}}$n, and Shnirman}}]{mak01}
\bibinfo{author}{\bibfnamefont{Y.}~\bibnamefont{Makhlin}},
  \bibinfo{author}{\bibfnamefont{G.}~\bibnamefont{Sch$\ddot{\mathrm{o}}$n}},
  \bibnamefont{and} \bibinfo{author}{\bibfnamefont{A.}~\bibnamefont{Shnirman}},
  \bibinfo{journal}{Rev.~Mod.~Phys.} \textbf{\bibinfo{volume}{73}},
  \bibinfo{pages}{357} (\bibinfo{year}{2001}).

\bibitem[{\citenamefont{Hassel and Sepp$\ddot{\mathrm{a}}$}(1999)}]{has99}
\bibinfo{author}{\bibfnamefont{J.}~\bibnamefont{Hassel}} \bibnamefont{and}
  \bibinfo{author}{\bibfnamefont{H.}~\bibnamefont{Sepp$\ddot{\mathrm{a}}$}}
  (\bibinfo{year}{1999}), \bibinfo{note}{in Proc. of 22$^{\mathrm{nd}}$ Int.
  Conf. on Low Temp. Phys.}

\bibitem[{\citenamefont{Zhu and Wang}(2002)}]{zhu02}
\bibinfo{author}{\bibfnamefont{S.-L.} \bibnamefont{Zhu}} \bibnamefont{and}
  \bibinfo{author}{\bibfnamefont{Z.}~\bibnamefont{Wang}},
  \bibinfo{journal}{Phys.~Rev.~A} \textbf{\bibinfo{volume}{66}},
  \bibinfo{pages}{042322} (\bibinfo{year}{2002}).

\bibitem[{\citenamefont{Aunola et~al.}(2000)\citenamefont{Aunola, Toppari, and
  Pekola}}]{aun00}
\bibinfo{author}{\bibfnamefont{M.}~\bibnamefont{Aunola}},
  \bibinfo{author}{\bibfnamefont{J.~J.} \bibnamefont{Toppari}},
  \bibnamefont{and} \bibinfo{author}{\bibfnamefont{J.~P.}
  \bibnamefont{Pekola}}, \bibinfo{journal}{Phys.~Rev.~B}
  \textbf{\bibinfo{volume}{62}}, \bibinfo{pages}{1296} (\bibinfo{year}{2000}).

\bibitem[{\citenamefont{Aunola}(2001)}]{aun01}
\bibinfo{author}{\bibfnamefont{M.}~\bibnamefont{Aunola}},
  \bibinfo{journal}{Phys.~Rev.~B} \textbf{\bibinfo{volume}{63}},
  \bibinfo{pages}{132508} (\bibinfo{year}{2001}).

\bibitem[{\citenamefont{Berry}(1984)}]{ber84}
\bibinfo{author}{\bibfnamefont{M.~V.} \bibnamefont{Berry}},
  \bibinfo{journal}{Proc. R. Soc. London, Ser. A}
  \textbf{\bibinfo{volume}{392}}, \bibinfo{pages}{45} (\bibinfo{year}{1984}).

\bibitem[{\citenamefont{Simon}(1983)}]{sim83}
\bibinfo{author}{\bibfnamefont{B.}~\bibnamefont{Simon}},
  \bibinfo{journal}{Phys. Rev. Lett.} \textbf{\bibinfo{volume}{51}},
  \bibinfo{pages}{2167} (\bibinfo{year}{1983}).

\bibitem[{\citenamefont{Nakahara}(1990)}]{nak90}
\bibinfo{author}{\bibfnamefont{M.}~\bibnamefont{Nakahara}},
  \emph{\bibinfo{title}{Geometry, topology, and physics}}
  (\bibinfo{publisher}{IOP Publishing}, \bibinfo{address}{Bristol, New York},
  \bibinfo{year}{1990}), pp. \bibinfo{pages}{29--30, 364--372}.

\bibitem[{\citenamefont{Chang and Simon}(1995)}]{cha95}
\bibinfo{author}{\bibfnamefont{M.-C.} \bibnamefont{Chang}} \bibnamefont{and}
  \bibinfo{author}{\bibfnamefont{Q.~N.} \bibnamefont{Simon}},
  \bibinfo{journal}{Phys. Rev. Lett.} \textbf{\bibinfo{volume}{75}},
  \bibinfo{pages}{1348} (\bibinfo{year}{1995}).

\bibitem[{\citenamefont{Rudin}(1973)}]{rud73}
\bibinfo{author}{\bibfnamefont{W.}~\bibnamefont{Rudin}},
  \emph{\bibinfo{title}{Functional analysis}}
  (\bibinfo{publisher}{McGraw-Hill}, \bibinfo{address}{New York},
  \bibinfo{year}{1973}).

\bibitem[{\citenamefont{Ingold and Nazarov}(1992)}]{ing92}
\bibinfo{author}{\bibfnamefont{G.-L.} \bibnamefont{Ingold}} \bibnamefont{and}
  \bibinfo{author}{\bibfnamefont{Y.~V.} \bibnamefont{Nazarov}}, in
  \emph{\bibinfo{booktitle}{Single charge tunnelling, Coulomb blockade
  phenomena in nanostructures}}, edited by
  \bibinfo{editor}{\bibfnamefont{H.}~\bibnamefont{Grabert}} \bibnamefont{and}
  \bibinfo{editor}{\bibfnamefont{M.~L.} \bibnamefont{Devoret}}
  (\bibinfo{publisher}{Plenum}, \bibinfo{address}{New York},
  \bibinfo{year}{1992}), chap.~\bibinfo{chapter}{2}.

\bibitem[{\citenamefont{Tinkham}(1996)}]{tink96}
\bibinfo{author}{\bibfnamefont{M.}~\bibnamefont{Tinkham}},
  \emph{\bibinfo{title}{Introduction to superconductivity, 2$^{\mathrm{nd}}$
  ed.}} (\bibinfo{publisher}{McGraw-Hill}, \bibinfo{address}{New York},
  \bibinfo{year}{1996}), pp. \bibinfo{pages}{257--277}.

\bibitem[{\citenamefont{Aunola}(2003)}]{aun02d}
\bibinfo{author}{\bibfnamefont{M.}~\bibnamefont{Aunola}}, \bibinfo{journal}{J.
  Math. Phys.} \textbf{\bibinfo{volume}{44}}, \bibinfo{pages}{1913}
  (\bibinfo{year}{2003}).

\bibitem[{int()}]{integr}
\bibinfo{note}{For those not familiar with differential geometry, it suffices
  to state that this is just a generalisation of two-dimensional integral onto
  curved manifolds and that for a flat surface $dx\wedge dy=-dy\wedge dx$
  reduces to the conventional $dxdy$.}

\bibitem[{\citenamefont{Caldeira and Leggett}(1983)}]{cald83}
\bibinfo{author}{\bibfnamefont{A.}~\bibnamefont{Caldeira}} \bibnamefont{and}
  \bibinfo{author}{\bibfnamefont{A.}~\bibnamefont{Leggett}},
  \bibinfo{journal}{Ann. Phys.} \textbf{\bibinfo{volume}{149}},
  \bibinfo{pages}{374} (\bibinfo{year}{1983}).

\bibitem[{\citenamefont{Cottet et~al.}(2001)\citenamefont{Cottet, Steinbach,
  Joyez, Vion, Pothier, Esteve, and Huber}}]{cottet2000}
\bibinfo{author}{\bibfnamefont{A.}~\bibnamefont{Cottet}},
  \bibinfo{author}{\bibfnamefont{A.}~\bibnamefont{Steinbach}},
  \bibinfo{author}{\bibfnamefont{P.}~\bibnamefont{Joyez}},
  \bibinfo{author}{\bibfnamefont{D.}~\bibnamefont{Vion}},
  \bibinfo{author}{\bibfnamefont{H.}~\bibnamefont{Pothier}},
  \bibinfo{author}{\bibfnamefont{D.}~\bibnamefont{Esteve}}, \bibnamefont{and}
  \bibinfo{author}{\bibfnamefont{M.}~\bibnamefont{Huber}}, in
  \emph{\bibinfo{booktitle}{Macroscopic Quantum Coherence and Computing}},
  edited by \bibinfo{editor}{\bibfnamefont{D.}~\bibnamefont{Averin}},
  \bibinfo{editor}{\bibfnamefont{B.}~\bibnamefont{Ruggiero}}, \bibnamefont{and}
  \bibinfo{editor}{\bibfnamefont{P.}~\bibnamefont{Silvestrini}},
  \bibinfo{organization}{Macroscopic Quantum Coherence 2}
  (\bibinfo{publisher}{Plenum Publishers}, \bibinfo{address}{New York},
  \bibinfo{year}{2001}), vol. \bibinfo{volume}{145}.

\end{thebibliography}

\end{document}